\documentclass[twocolumn]{revtex4}
\usepackage{epsfig}

\begin{document}

\title{Influence of polymer excluded volume on the phase behavior of
colloid--polymer mixtures} \author{P.G. Bolhuis$^*$,
A. A. Louis$^\ddagger$, and J-P. Hansen$^\ddagger$}
\affiliation{$^*$Dept. of Chemical Engineering, University of
Amsterdam, Nieuwe Achtergracht 166, 1018 WV, Amsterdam, Netherlands\\
$^\ddagger$Dept. of Chemistry, University of Cambridge, Lensfield
Road, CB2 1EW, Cambridge, UK}

\begin{abstract}
 We determine the depletion-induced phase-behavior of hard sphere
colloids and interacting polymers by large-scale Monte Carlo
simulations using very accurate coarse-graining techniques.  A
comparison with standard Asakura-Oosawa model theories and simulations
shows that including excluded volume interactions between polymers
leads to qualitative differences in the phase diagrams. These effects
become increasingly important for larger relative polymer size. Our
simulations results agree quantitatively with recent experiments.
\end{abstract}
\maketitle

Adding a sufficient amount of non-adsorbing polymer chains to a stable
colloidal dispersion can cause a depletion-induced separation of the
dispersion into colloid-rich and colloid-poor phases, a striking
effect which has been extensively investigated
experimentally\cite{LiIn75,Cald93,Ilet95,Verh96,Bodn97,Verm98,Rent01,Rama02}
and which has important industrial\cite{Russ89} and
biological\cite{Zimm93} implications.  The entropy-driven depletion
attraction between colloidal particles can be tuned by varying the
polymer-to-colloid size ratio $q=R_g/R_c$ (where $R_g$ is the polymer
radius of gyration and $R_c$ the radius of the spherical colloids) and
the polymer concentration, thus providing a unique opportunity of
generating rich phase behavior.  A theoretical description of
colloid-polymer mixtures is a challenging Statistical Mechanical
problem because of the large length scale differences between the size
of the colloids and the polymer segments. The simplest and most widely
made assumption is to consider ideal (non-interacting) polymers
obeying Gaussian or random walk statistics\cite{deGe79}.  A further
simplification is provided by the Asakura-Oosawa (AO) model\cite{Asak58}
whereby polymer coils are treated as mutually penetrable spheres,
which are excluded from a sphere of radius $(R_c + R_g)$ around each
colloid. Gast et al.\cite{Gast83} and Lekkerkerker et al.\cite{Lekk92}
used this model to calculate the phase diagrams of colloid-polymer
systems.  Computer simulations of hard sphere (HS) colloids and ideal
lattice polymers\cite{Meij94} yielded results in good agreement with
predictions based on the simpler AO model. Thus the phase behavior of
mixtures of HS  colloids and ideal polymers is well
understood, at least for size ratios $q \leq 1$.  However, polymers
rarely behave as ideal, except perhaps near the theta
point\cite{deGe79}.  The more general problem involving {\em
interacting} polymer chains is much more difficult and over half a
century of theoretical work has shown that excluded volume
interactions between monomers lead to important qualitative and
quantitative differences in the properties of polymer
solutions\cite{deGe79}.  This letter presents the first large scale
systematic simulations of the phase behavior of mixtures of colloids
and interacting polymers for three size ratios $q$, and compares the
calculated phase diagrams to recent experimental
data\cite{Verh96,Bodn97,Rent01,Rama02}.

Some earlier attempts to account for polymer interactions in
colloid-polymer mixtures were based on a perturbation theory around
$\theta$-point conditions\cite{Warr95} or on integral
equations\cite{Fuch01}.  In the present work the self-avoiding walk
(SAW) model is adopted for the interacting polymers, which is known to
be a very good representation of polymers in good
solvent\cite{deGe79}.  A full scale simulation of $N_c$ colloidal hard
sphere particles and $N_p$ polymer chains, each made up of $L \gg 1$
monomers or Kuhn segments would be a daunting task.  However,
large-scale simulations become feasible within a coarse-grained
description of the polymers, whereby the latter are represented as
single particles interacting via an effective pair potential between
their centers of mass (CM).  Such effective pair potentials can be
calculated by a well controlled tracing out of the individual monomer
degrees of freedom along the ``polymer as soft colloids'' approach we
have recently put forward\cite{Loui00,Bolh01,Bolh01a,Bolh02}.  This
approach was shown to reproduce, within statistical simulation
uncertainties, the correct equation of state of pure interacting
polymer solutions\cite{Bolh01,Bolh02}, as well as the correct one-body
free energy of inserting a single colloid into a polymer solution, and
the related polymer surface tension\cite{Loui02}.  Compared to very
time-consuming monomer-level simulations, the coarse-graining method
moreover yields accurate depletion potentials between two
plates\cite{Bolh01} or between two colloidal spheres\cite{Loui02a} for
polymer volume fractions up to $\phi_p \equiv \frac{4}{3}\pi R_g^3
N_p/V \approx 2$.
The success with the
two-colloid problem suggests that the coarse-graining procedure may be
fruitfully extended to the full many-body problem of the phase
behavior of polymer-colloid mixtures as long as $\phi_p$ is not
much greater than $1$ (which marks the cross-over to the semi-dilute regime).
\begin{figure*}
\epsfig{figure= 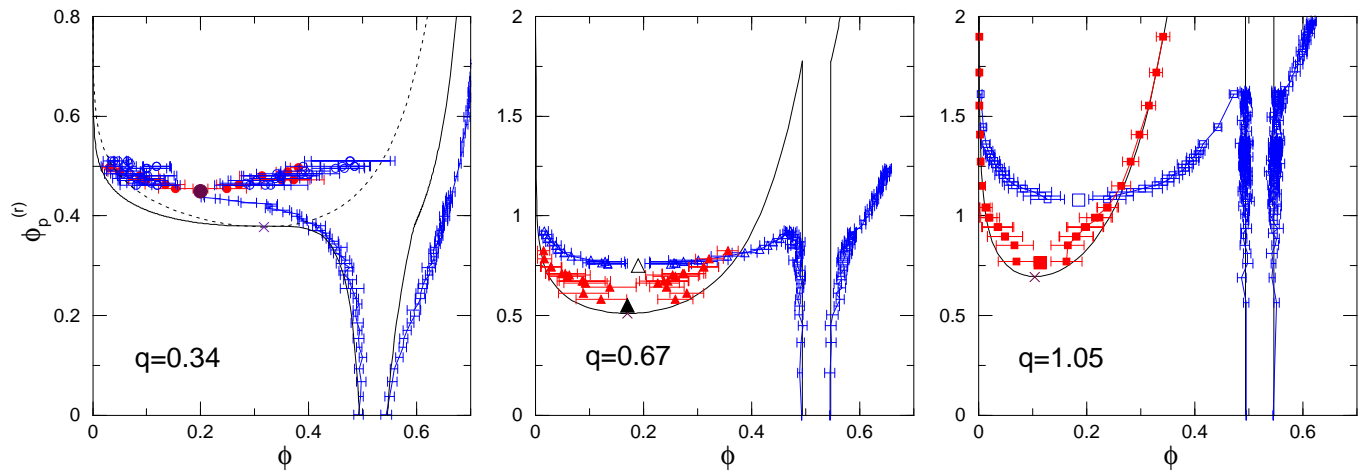,width=18cm}
\caption{\label{fig1} Phase diagrams in the polymer reservoir
concentration--colloid volume fraction representation.  The open
symbols denote the fluid-fluid and the dashed lines with error bars
the fluid-solid coexistence curves for interacting polymers.  The filled symbols
denote the fluid-fluid phase-lines for the two-component AO model. The
critical points are indicated by a larger open and closed symbol for
the interacting polymers and AO model, respectively.  The thin solid
curves represent free-volume perturbation theory phase diagram, with a
cross denoting the location of the critical point. The dashed line in
the left panel is the free volume theoretical metastable fluid-fluid
binodal.}
\end{figure*}

Firstly, as described in our previous publications, the effective
potentials for the polymer-polymer interaction were obtained from
simulations of a bulk system of self avoiding walks (SAW), at various
concentrations.
All simulations were for polymer chains of length $L=500$ segments,
with the zero-concentration $R_g=16.83$ lattice units. This length is
sufficient to show proper scaling behavior in the semidilute
regime~\cite{Bolh01}.  The concentration-dependent effective
interactions $v_{pp}(r;\phi_p)$ were obtained by inversion of the CM
radial distribution function $g(r)$ using the hypernetted chain (HNC)
integral equation\cite{Bolh01}, and were accurately parameterized by sums
of Gaussians~\cite{Bolh02}.

The concentration-dependent potentials $v_{cp}(r;\phi_p)$ for the
colloid-polymer interaction at each $q$ and $\phi_p$ were obtained
from simulations of a single hard sphere in a solution of SAW
polymers.  The CM concentration profile was inverted by using a
two-component version of the HNC equation, and the resulting
$v_{cp}(r;\phi_p)$ was fitted to an exponential form~\cite{Bolh02}. 

The direct colloid-colloid interaction $v_{cc}(r)$ was taken to be
hard-sphere like.  These three interparticle potentials are the basis
of the coarse-graining scheme.  Each polymer is now reduced to a
single effective particle, opening the way to large-scale simulations
of a binary mixture of polymers and colloids.  It is important to note
that in a system with a finite density of colloids, the polymer
concentration parameter $\phi_p$ in the effective potentials
$v_{pp}(r;\phi_p)$ and $v_{cp}(r;\phi_p)$ must be chosen to be that in
a reservoir of a pure polymer system in {\em osmotic equilibrium} with
the two-component system of interest.  In other words, as discussed in
detail by other authors\cite{Lekk92,Dijk98,Dzub01}, the effective
interactions should be taken at the chemical potential $\mu_p$ of the
polymers in an osmotic reservoir.  

The Gibbs-Ensemble Monte Carlo (GEMC)
technique\cite{Pana87,frenkelbook} is naturally suited for studying
fluid-fluid phase separation.  The chemical potential $\mu_p$ was
fixed by a standard grand canonical prescription~\cite{Bolh94} and the
number of colloidal particles were fixed at $N_c=108, N_c=150$ and
$N_c=200$ for polymer-colloid size ratios of $q = 0.34, q=0.67$ and
$q=1.05$ respectively.  The GEMC simulations yielded histograms for
the polymer and colloid densities in both boxes. At chemical
potentials above a critical value, the two boxes show different colloid
densities, corresponding to phase separation.

Besides the fluid-fluid phase separation, colloid crystallization can
also occur in colloid-polymer mixtures. At zero polymer concentration
the fluid-solid coexistence occurs at colloid volume fractions of
$\phi_c=0.494$ and $\phi_c=0.545$, as expected for a pure HS
system. The effect of the added polymer is initially to widen the
coexistence curve. At very high polymer concentration, a dense
colloidal crystal can be in equilibrium with a very dilute colloidal
``gas'' (see e.g. Ref.\cite{Loui01a}).  Experiments and previous
approximate theoretical work show that if $q \lesssim 0.3$ the
fluid-fluid critical point becomes metastable w.r.t.\ the
crystallization phase-line.  For this reason, it is important to also
calculate the fluid-solid phase lines, which was done using Kofke's
Gibbs-Duhem integration technique~\cite{Kofk93,frenkelbook}.  Starting
with zero polymer activity at HS fluid-solid coexistence, we performed
a series of $N\mu_p PT$ ensemble simulations, integrating the
Clausius-Clapeyron equation $dP/d\mu_p = \Delta N_p / \Delta V$ along
the coexistence line, where $\Delta N_p$ and $\Delta V$ are the
differences in number of polymers and in volume between the two phases
respectively.

To compare with ideal polymer theories, full GEMC simulations were
also performed for the AO penetrable sphere model of
polymers\cite{Asak58}, with the same numbers of colloids and
size-ratios as for the interacting polymer systems.  These simulations
should provide an accurate representation of a true ideal polymer
system\cite{Meij94}.  For comparison, we also calculated the
phase-diagrams within the free-volume theory of Lekkerkerker {\em et
al.}\cite{Lekk92}.

The complete phase diagrams for the ideal and interacting polymer
models at all three size ratio's are plotted in Fig.~\ref{fig1}, in
the polymer reservoir concentration-colloid volume fraction
representation.  Although the polymer chemical potential is the
natural control variable in the simulations, the reservoir polymer
concentration $\phi_p^{(r)}$, directly obtainable from $\mu_p$ through the SAW
equation of state\cite{Bolh01,Bolh02}, is a more suitable variable
because this brings the interacting and non-interacting polymer
phase-lines much closer to each other than a direct comparison of the
chemical potentials would.

 For $q=0.34$ the fluid-fluid binodal has just become metastable.
This result is consistent with experiments\cite{Ilet95}, and close to
the prediction of the AO model and other non-interacting polymer
theories\cite{Lekk92,Meij94}.  The direct AO model simulations show a
fairly good agreement with the interacting polymer simulations and are
in good qualitative agreement with the free-volume perturbation
theory, although the latter overestimates the critical colloid
density.  Although we did not explicitly calculate the triple point
for the two-component AO system, it is expected to be fairly well
located by the simpler free volume theory shown in the
figure\cite{Lekk92,Meij94}.  For $q=0.67$ and $q=1.05$ larger
differences are seen between the interacting polymers and the
non-interacting AO model system. The triple point predicted by the
latter moves to much higher polymer reservoir concentration than that
found for interacting polymers.  Whereas the critical point for
interacting polymers stays near $\phi_c = 0.2$, the ideal polymer
critical point moves to lower colloid densities, an effect that was
predicted in the literature\cite{Sear01}.  Finally we point out that
since the polymer reservoir concentration stays well below the
concentration $\phi_p=2$, i.e.\ in the regime where we previously
found good agreement for the two-body problem, we expect our
coarse-graining model to provide a very accurate representation of the
fully interacting polymer-colloid system.  Should the critical polymer
concentration continue to rise with increasing $q$, then the
coarse-graining method would become less trustworthy for large size
ratios.

\begin{figure}
\epsfig{figure= 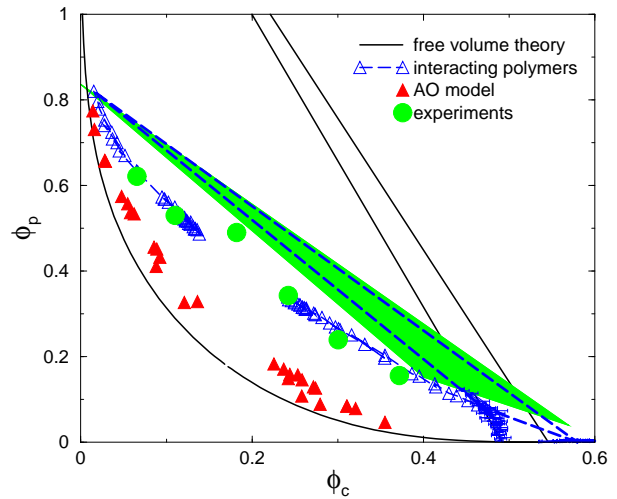,width=8cm,angle=0}
\caption{\label{fig2} Phase diagram in the polymer volume
fraction-colloid volume fraction representation for $q=0.67$. Symbols
as in Fig 1.  The large triangular areas denote the estimates for the
triple point: the shaded area corresponds to the experimental data
[7]. The simulated binodal agrees quantitatively with the experimental data
of ref.~\protect\cite{Rama02}}.
\end{figure}

Experiments are usually done at fixed polymer concentration, so that
the tie-lines are no longer horizontal.  This representation is shown
in Fig.\ref{fig2} for $q=0.67$.  Triple points turn into triangular
areas in which three phases coexist. These have been measured
experimentally using a mixture of PMMA-latex and
polystyrene\cite{Rent01}.  We compare the experimental triple point
results for $q=0.57$ to the theoretical $q=0.67$ diagram. Although the
systems are not entirely equivalent, a much better agreement is found
with the interacting polymer simulations than with the AO model
estimates.  The simulations also agree quantitatively with the
fluid-fluid binodal which was accurately measured for $q=0.677$ in
recent experiments on silica particles in toluene\cite{Rama02}.
Similar quantitative agreement (not shown) was also found between the $q=0.34$
simulations and $q=0.377$ experimental binodals\cite{Rama02}.

\begin{figure}[t]
\epsfig{figure= 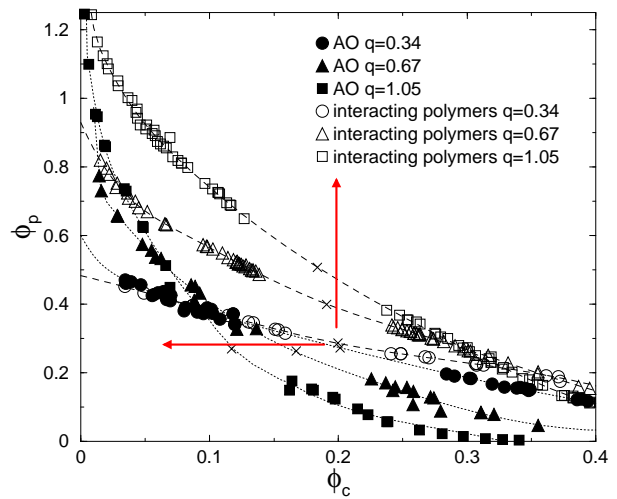,width=8cm,angle=0}
\caption{\label{fig3} Fluid-fluid binodals in the polymer
volume fraction-colloid volume fraction plane. Symbols as in Fig 1. The crosses
indicate the estimated position of the critical point. Note the
qualitative differences in the effect of increasing $q$ on the
critical points as indicated by the arrows. }
\end{figure}

It is also instructive to compare the fluid-fluid binodals for
different size-ratios, as done in Fig.~\ref{fig3}.  For small size
ratio the AO and the interacting polymer estimates are close, as also
seen in Fig.~\ref{fig1}.  This is not surprising, since we have shown
previously that a cancellation of errors in the AO model leads to
fairly good effective pair potentials for small $q$ and
$\phi_p$\cite{Loui02a}. But the simulations for $q>0.34$ exhibit
qualitatively different behavior: the critical colloid density of the
AO model decreases whereas the polymer concentration does not change
much.  Including polymer interactions has the opposite effect: the
critical polymer concentration increases and the colloid density
hardly changes.  Recent integral equation
calculations\cite{Rama02,Fuch01} of the spinodal phase lines also show
an increase in the polymer concentration with increasing $q$.

\begin{figure}
\epsfig{figure= 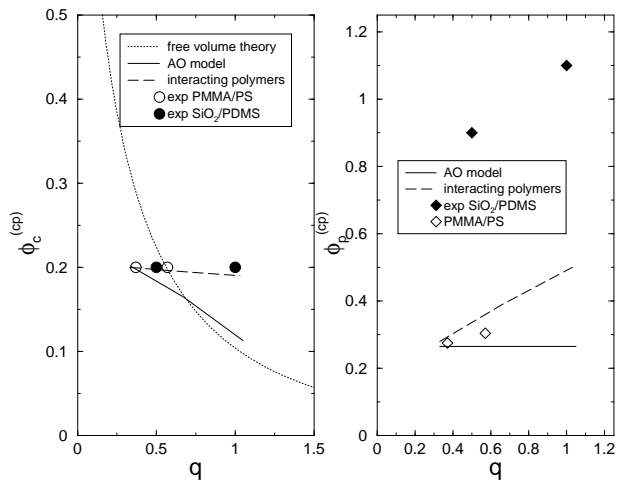,width=8cm,angle=0}
\caption{\label{fig4} The behavior of the critical concentrations as a
function of size ratio $q$. The interacting polymer predictions for
the colloid critical volume fraction $\phi_c^{(cp)}$ agree with
experimentally determined critical concentrations. The free volume
theory overestimates $\phi_c^{(cp)}$ for small size ratio but
underestimates it for large size ratio.  The critical polymer volume
fractions $\phi_p^{(cp)}$ for the interacting polymer simulations are
in reasonable agreement with the experiments of
Ref.~\protect\cite{Ilet95,Rent01} but compare less favorably to those
of Refs~\protect\cite{Bodn97} and~\cite{Verh96}.}
\end{figure}

The behavior of the critical point is summarized in Fig.~\ref{fig4},
where the predicted trend for the critical colloid density is shown to
agree very well with another set of
experiments\cite{Verh96,Bodn97}. We have recently performed direct
simulations for much larger size ratios $(q > 5)$, and found that the
colloid critical point is near the same density as found
here\cite{largepols}.  This suggests that even in the so called
``protein limit'', $q\rightarrow \infty$, the colloid critical density
is finite. The situation for the critical polymer volume fraction is
less clear. The present simulations seem to agree fairly well with the
experimental estimates from Ref.~\cite{Ilet95,Rent01} although we
should stress that the critical point was not determined
accurately. Two other experimental results from Ref.~\cite{Verh96} and
Ref.~\cite{Bodn97} based on mixtures of silica and PDMS polymer are
also included. These studies determined the critical point accurately,
and the colloid density agrees well with our simulations. However, the
polymer concentrations are twice as large, which might be due to the
highly polydisperse polymers used by these authors ($M_w/M_n=2$,
versus $M_w/M_n=1.04$ in Ref~\cite{Rent01}).

In summary, this letter presents the first large scale computer
simulations of the full equilibrium phase-diagram of a binary mixture
of colloids and interacting polymers. These simulations would not have
been possible without using the accurate ``polymers as soft colloids''
coarse-graining approach.  Even though the phase-separation happens
mainly in the dilute regime of the polymer solution, we find important
qualitative differences with ideal polymer behavior. In particular,
the absolute polymer concentration $\phi_p$ at the critical point
increases with increasing $q$ for interacting polymers, while it
decreases with increasing $q$ for non-interacting polymers.
Similarly, the critical colloid packing fraction $\phi_c^{(cp)}$
remains nearly constant for interacting polymers, while it decreases
with increasing $q$ for non-interacting polymers.  We also showed that
by including excluded volume interactions we find quantitative
agreement with experiments.  In conclusion then, just as excluded
volume interactions are known to strongly influence the behavior of
polymer solutions, we find here that polymer excluded volume
interactions have important qualitative effects on the behavior of
polymer-colloid mixtures.

AAL acknowledges support from the Isaac Newton Trust, Cambridge. We
thank C.F. Zukoski and S. Ramakrishnan for sending a copy of their
data.


\end{document}